\documentclass[conference]{IEEEtran}
\IEEEoverridecommandlockouts
\usepackage{cite}
\usepackage{amsmath,amssymb,amsfonts}
\usepackage{algorithmic}
\usepackage{graphicx}
\usepackage{textcomp}
\usepackage{xcolor}
\usepackage{booktabs}
\usepackage{balance}
\usepackage{hyperref}
\usepackage{comment}
\usepackage{eso-pic}
\def\BibTeX{{\rm B\kern-.05em{\sc i\kern-.025em b}\kern-.08em
    T\kern-.1667em\lower.7ex\hbox{E}\kern-.125emX}}
\begin{document}

\title{Forecasting Individual NetFlows using a Predictive Masked Graph Autoencoder}

\author{\IEEEauthorblockN{Georgios Anyfantis}
\IEEEauthorblockA{\textit{Department of Computer Architecture} \\
\textit{Universitat Politècnica de Catalunya}\\
Barcelona, Spain \\
georgios.anyfantis@upc.edu}
\and
\IEEEauthorblockN{Pere Barlet-Ros}
\IEEEauthorblockA{\textit{Department of Computer Architecture} \\
\textit{Universitat Politècnica de Catalunya}\\
Barcelona, Spain \\
pere.barlet@upc.edu}
}

\maketitle

\AddToShipoutPictureFG*{%
  \AtPageLowerLeft{%
    \put(54,28){%
      \parbox{0.85\paperwidth}{%
        \scriptsize
        \textcopyright~2026 IEEE. Personal use of this material is permitted.
        Permission from IEEE must be obtained for all other uses.
      }%
    }%
  }%
}

\begin{abstract}
In this paper, we propose a proof-of-concept Graph Neural Network model that can successfully predict network flow-level traffic (NetFlow) by accurately modelling the graph structure and the connection features. We use sliding-windows to split the network traffic in equal-sized heterogeneous bidirectional graphs containing IP, Port, and Connection nodes. We then use the GNN to model the evolution of the graph structure and the connection features. Our approach shows superior results when identifying the Port and IP to which connections attach, while feature reconstruction remains competitive with strong forecasting baselines. Overall, our work showcases the use of GNNs for per-flow NetFlow prediction.
\end{abstract}


\section{Introduction}

Forecasting network traffic can be used for traffic engineering, routing, improving resource allocation, and service orchestration by anticipating future patterns that are not visible at the aggregate level \cite{Pf_lb_2019}.

A substantial body of research on the prediction of network traffic focuses on aggregated metrics \cite{10015152}, such as link utilisation over time. However, there has been limited research on predicting individual NetFlows \cite{jahnke2018finegrainednetworkflow}. This is not surprising due to the increased complexity of the forecasting problem \cite{jahnke2018finegrainednetworkflow}. Flows tend to arrive at large volumes, combine both numerical and categorical data, and exhibit periodic changes, making them very difficult to model correctly. Moreover, early work on fine-grained prediction noted that per-flow forecasting was impractical \cite{Pf_lb_2019}.

However, per-flow forecasting can be very valuable as it preserves a lot of information that aggregated forecasting removed \cite{Pf_lb_2019}. This can help predict potential bottlenecks and resource demand, support more fine-grained network engineering decisions, and improve the modelling of distributed applications. Finally, recent work on flow-level simulation indicates that aggregate models are insufficient for tasks that require fine-grained information \cite{li2025m4learnedflowlevelnetwork}.

Previous work on ML based approaches for network traffic has suffered from the smoothing problem on aggregated metrics as aggregated smoothing tends to hide micro-bursts \cite{draft-irtf-nmrg-ai-challenges}. These micro-bursts can contain important information about the network that is lost.

Graph Neural Networks (GNNs) have been found to be able to solve the stagnation problem observed when modelling networks by leveraging the graph structure \cite{bdcc9110270}. This is not possible in traditional models, as they cannot leverage the relationship between individual flows. GNNs are Neural Networks that use both the node features and the graph structure to capture relational information \cite{zhang2019graph}.

Per-flow forecasting is an inherently relational task. A NetFlow is not an isolated observation, but rather a part of a larger communication pattern \cite{XU2024110495}. As such, modelling NetFlows solely as tabular data may miss important relational information \cite{pujolperich2021unveilingpotentialgraphneural}. GNNs are ideal for this task, as they capture the network interactions and better model the changing nature of the network, making GNNs the ideal choice for per-flow prediction.

IPFIX and NetFlows remain a widely used and practical format to capture network telemetry for monitoring. They provide a standardised representation of communication behaviour inside the network without requiring full payload inspection \cite{Aitken2013-xz}. This makes them an ideal format for use for forecasting tasks.

In this paper, we employ Graph Neural Network (GNN) for fine-grained per-flow prediction using the NetFlow format. Our work is a proof-of-concept approach showcasing the capability of GNNs for per-flow prediction. We compare our proposed solution with other strong Machine Learning forecasting methodologies. Our results indicate that GNNs are very well suited for per-flow prediction and showcase a clear structural advantage and competitive feature reconstruction compared to the baselines employed in this paper.

\section{Problem Description}
\label{sec:related-work}

Flow prediction is usually tackled as a time-series problem \cite{10015152}. In this paper, we are looking at predicting flows as a sliding window one-step forecasting problem. One-step forecasting problems look at taking the current step and predicting the next step without any input from the previous steps. In our approach, we are taking the current step that represents the network traffic and predict the next step with the future network traffic.

In the per-flow prediction, we slice the traffic into non-overlapping windows of equal lengths. Then we feed the current window to the model and we try to predict the next sequence of flows.

To evaluate the efficacy of GNNs for per-flow prediction, we have implemented some strong sequence-prediction backbones. These methodologies share the same encoder and decoder architecture, but we switch the backbone to implement each architecture.

The Long-Short-Term Memory (LSTM) model \cite{6795963} is a well-known recurrent model architecture that is widely used for forecasting and modelling \cite{LINDEMANN2021650}. The architecture works by ingestion of the input sequentially and updating its internal hidden state through gated updates, helping the model learn temporal dependencies over long horizons \cite{6795963}. LSTMs have been extensively researched and are known as a standard forecasting architecture \cite{LINDEMANN2021650}.

Temporal Convolution Network (TCN) is a forecasting architecture that switches recurrence with causal one-dimensional convolutions \cite{bai2018empiricalevaluationgenericconvolutional}. In their standard form, TCNs use causal convolutions with dilation and residual connections, allowing the receptive field to grow while preserving temporal order. This makes them an attractive alternative to standard recurrent architectures for modelling long-range dependencies. In the original paper, TCN was found to perform better than traditional approaches in long-range dependency modelling \cite{bai2018empiricalevaluationgenericconvolutional} and a recent survey indicates that TCN is widely used in time-series forecasting \cite{Kong2025-bz}.

Transformers are attention-based models that were originally built as encoder-decoder architectures without recurrence or convolutions \cite{DBLP:journals/corr/VaswaniSPUJGKP17}. The Transformer's main strength is the use of the self-attention mechanism for capturing and modelling long-range dependencies. In time series forecasting, Transformer based models have been proposed and have been successfully applied to time series forecasting and tasks \cite{wen2023transformerstimeseriessurvey}.

DLinear is a simple linear forecasting architecture \cite{zeng2022transformerseffectivetimeseries}. It works by decomposing the input into a trend component and a seasonal or remainder component using a moving-average decomposition. Then it applies separate one-layer linear projections to the two components and combines them to a final projection. It is a more recent model than LSTM \cite{6795963}, TCN \cite{bai2018empiricalevaluationgenericconvolutional} and Transformers \cite{DBLP:journals/corr/VaswaniSPUJGKP17} it has been found to be a very good baseline used for benchmarking in long-term time-series forecasting.

Graph Neural Networks are a class of Neural Networks that uses structured graph data to learn representations by propagating and aggregating information across connected graph nodes \cite{4700287}. GraphSAGE is one of the most well-known GNN algorithms used in inductive formulations, and it works by aggregating features from local neighbourhoods, allowing the model to capture both the structure and the attributes of the graph \cite{hamilton2018inductiverepresentationlearninglarge}.

\begin{figure*}[!t]
\centering
\includegraphics[width=0.8\linewidth]{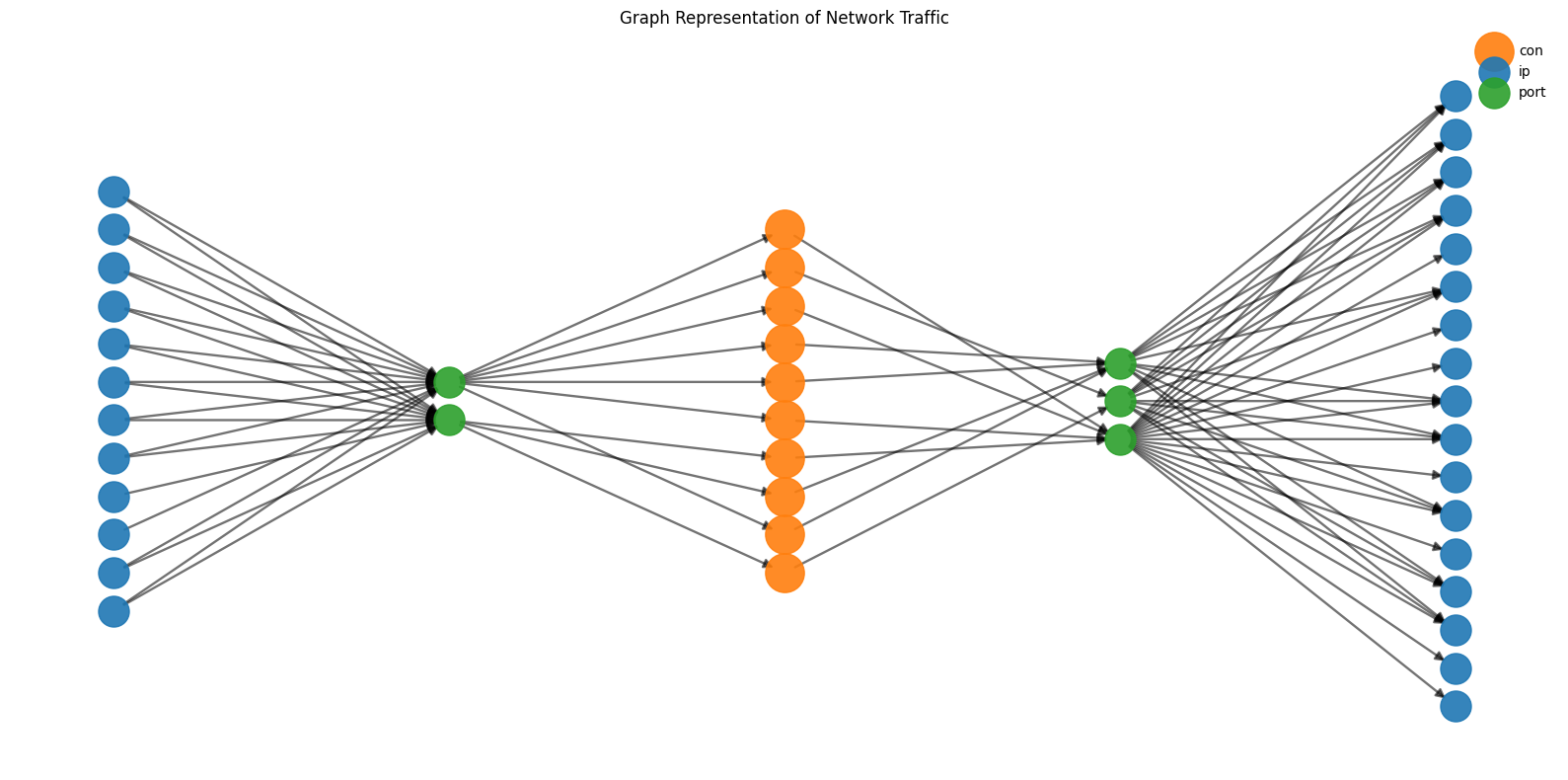}
\caption{A graph representation on how the NetFlows are represented in our Graphs. This is a sampled subset of 10 connections.}
    \label{fig:graph-rep}
\end{figure*}

\begin{figure*}[!t]
\centering
\includegraphics[width=0.9\linewidth]{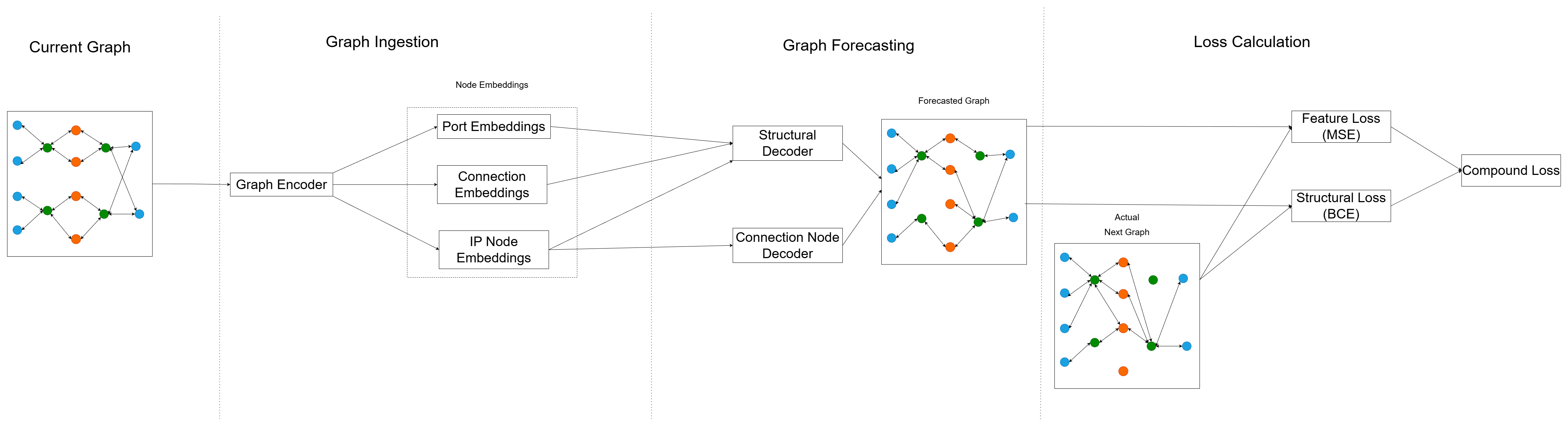}
\caption{A representation of our architecture and its training procedure.}
    \label{fig:autoencoder}
\end{figure*}

\section{Methodology}

\subsection{Dataset and Graph Creation}

For this paper, we used the UNSW-NB15 dataset \cite{7348942}. We chose to use this dataset as it is a widely used dataset with individual flows. For this paper, we chose to extract NetFlows V9 \cite{claise2004cisco} from raw PCAP files. For extraction, we used NFStream \cite{AOUINI2022108719}, which is a lightweight Python library that can extract NetFlows from raw PCAP files. We followed a similar pre-processing and graph creation approach as we had followed in our previous paper \cite{anyfantis2026autographadunsupervisednetworkanomaly}.

For data pre-processing, we use L2 normalisation for numerical data. We chose this normalisation as it is very efficient and avoids the requirement of pre-training an encoder and having to re-tune to account for distribution drifts. For categorical data, such as ports and protocols, we use One-Hot Encoding (OHE).

For the OHE, we used pre-determined values for the most important categories. The schema can be seen in Table \ref{tab:ohe_schema}. Due to the nature of the UNSW-NB15 dataset, we discarded the IP version since all traffic is IPv4.

\begin{table}[h]
\centering
\caption{One-Hot Encoding (OHE) Mapping Schema}
\label{tab:ohe_schema}
\begin{tabular}{@{}lll@{}}
\toprule
\textbf{Feature} & \textbf{OHE Category} & \textbf{Original Values / Logic} \\ \midrule
\textbf{Ports} & Port 0 & 0 \\
 & Port 53 & 53 \\
 & Port 123 & 123 \\
 & Port 443 & 443 \\
 & Well-Known & $0 \le x \le 1023$ (Excl. specific ports) \\
 & Registered & $1024 \le x \le 49151$ \\
 & Dynamic/Private & $49152 \le x \le 65535$ \\ \midrule
\textbf{Protocols} & ICMP & 1 \\
 & TCP & 6 \\
 & UDP & 17 \\
 & Others & All other protocol values \\ \midrule
\end{tabular}
\end{table}

Regarding graph building, we chose to use fixed sized sliding windows. The data is split into non-overlapping windows of length 512. We chose this length because it has been well documented that the baseline methods can handle this sequence length very well. The graphs were constructed as heterogeneous. We are using IP nodes that have placeholder features initialised to ones, 8 unique Port nodes with initialised placeholder features to ones, and the connection nodes with connection features. The port nodes represent the OHE categories, and their features were initialised to ones as there was no performance enhancement when richer port features were used. A representation of how NetFlows are represented in our graphs can be found in Figure \ref{fig:graph-rep}.

For the baselines, we have followed the same pre-processing and sliding window splitting. The IPs and Ports are mapped to unique IDs that the model then processes. The ports are mapped using the OHE categories.

The whole pre-processing step was implemented using NumPy \cite{harris2020array}.

In terms of dataset splits, we chose to allocate 70\% of the dataset in the training set, 20\% of the dataset in the validation set, and 10\% for the testing part. We used \textit{train\_test\_split} from Scikit-Learn \cite{scikit-learn} to create the splits. The splits are random and do not use any of the dataset labels. Since only the current window is used for the prediction, we decided that this approach is appropriate since there is no temporal overlap in the training as each step is seen individually and no past steps are fed into the model. Although temporal split provides a stricter temporal generalisation and distribution drift due to the nature of our split and that this work is a proof-of-concept, we believe this approach is appropriate. 

\subsection{Model Architecture}

Our proposed model has been inspired by our previous work on AutoGraphAD \cite{anyfantis2026autographadunsupervisednetworkanomaly}. The model works by mapping the current graph into the future graph. For this task, we employ a Graph Autoencoder \cite{kipf2016variationalgraphautoencoders}.

For our proposed model, we have built on our previous work of using masked Graph Variational Autoencoders (GVAE) \cite{anyfantis2026autographadunsupervisednetworkanomaly}. Our architecture can be seen in Figure \ref{fig:autoencoder}.

For this paper, we are using a simple Graph Autoencoder (GAE) that adapts the Graph Masked Autoencoder Approach (GraphMAE) \cite{GraphMAE} and heterogeneous graph mask autoencoders (HGMAE) \cite{tian2023heterogeneousgraphmaskedautoencoders}. GraphMAE uses a learnable embedding to perform masking, similar to BERT \cite{devlin2019bertpretrainingdeepbidirectional}. We are using the same masking technique as proposed by HGMAE and GraphMAE but we do not perform any masking on the edges. We are focusing on reconstructing the masked nodes and all the graph edges.

We are using random edge dropping as part of the training to force the model to learn non-trivial solutions and better capture the structural information of the graph \cite{tian2023heterogeneousgraphmaskedautoencoders}. Thus, allowing for better reconstruction.

Our autoencoder uses GraphSAGE \cite{hamilton2018inductiverepresentationlearninglarge} as it is a proven algorithm that works well with unseen graphs.

For our node feature decoder, we are using a simple Multi-Layer Perceptron (MLP). We have chosen to use this architecture because it is simple and has proven to be very successful when used for GAE \cite{tian2023heterogeneousgraphmaskedautoencoders}. Moreover, the use of an MLP compared to a GNN decoder removes the need of needing the future graph's edge index, something that should not be relied on in a forecasting problem.

For structural reconstruction, we are using an enhanced Dot Product \cite{anyfantis2026autographadunsupervisednetworkanomaly}. The enhanced Dot Product is based on the original Dot Product \cite{kipf2016variationalgraphautoencoders} but is enhanced by using learnable weights to better learn the structural information of the graph.

For our backpropagation loss, we use composite loss. It is defined by the weighted sum of the Feature Loss and the Structural Loss. The user can assign importance to the Structural Loss by setting the balancing hyperparameter $\alpha$. The loss can be seen in Equation \ref{loss_equation}.

\begin{equation}
\label{loss_equation}
    L_{Total} = \alpha * StructLoss + (1-\alpha)*FeatLoss
\end{equation}

Our new model differs from AutoGraphAD as it does not use a GVAE architecture and during training only reconstructs the masked features. Additionally, it does not employ masking on the edges as they are not static between the current and next graph. Furthermore, we use a weighted sum loss instead of the highly configurable loss used in AutoGraphAD. Finally, AutoGraphAD focuses on reconstructing the inputted graph, whereas our proposed GNN focuses on predicting the next graph. Our code can be found here \footnote{\hyperlink{https://github.com/georgeani/NetFlow-Forecasting-MGAE}{https://github.com/georgeani/NetFlow-Forecasting-MGAE}}.

\subsection{Baseline Architecture}

For our baseline architecture, we are using a traditional masked autoencoder \cite{He2021MaskedAA}. We have chosen this since it is the closest to our Masked GAE and, for modelling problems, Masked Autoencoders are ideal. For masking, for categorical data we use a separate token while for numerical features we use a learnable embedding similar to how BERT works \cite{devlin2019bertpretrainingdeepbidirectional}.

Our baselines use a unified backbone architecture, which means that all baselines share the encoder and decoder mechanism. For the encoder, we are using individual embedding layers \cite{mikolov2013distributedrepresentationswordsphrases} to map the IP address octets. The embedding layers use learnable tables to better map values into fixed sized vectors that are then combined with the rest of the input features to form the input sequence.

Additionally, all baselines share the same decoding architecture. For a decoder, we employ an MLP with two heads. One that reconstructs the IP and Port information, and another that reconstructs the rest of the flow features. This architecture mirrors the flow reconstruction approach that is employed by our GNN. Only the algorithm used for future forecasting is swapped out based on the baseline that we are using. As mentioned in Section \ref{sec:related-work}, we use DLinear \cite{zeng2022transformerseffectivetimeseries}, TCN \cite{bai2018empiricalevaluationgenericconvolutional}, LSTM \cite{6795963}, and Transformer \cite{DBLP:journals/corr/VaswaniSPUJGKP17} architectures.

\subsection{Libraries Used}

We implemented the Baselines using PyTorch \cite{NEURIPS2019_bdbca288} and PyTorch Geometric \cite{fey2019fastgraphrepresentationlearning} for GNN. We used PyTorch Lightning \cite{Falcon_PyTorch_Lightning_2019} to allow faster code development and better reusability. Torchmetrics \cite{Detlefsen2022} was used for the evaluation metrics.

To run our experiments, we used a server with an Nvidia RTX 3090 with 24 GB of VRAM. The CPU of the server is an AMD Ryzen 3950X, 16-Core Processor. The server came with 64 GB of RAM and its operating system was Ubuntu 22.04.4 LTS.

\section{Experimental Results}

\subsection{Hyperparameter Tuning}

To evaluate and find the optimal hyperparameters, we have employed Optuna \cite{akiba2019optunanextgenerationhyperparameteroptimization}. Optuna is a framework that allows for an efficient and informed hyperparameter space search. It provides the user with an array of sampling and pruning strategies that allow the framework to perform informed hyperparameter sampling using a user defined metric.

In our case, both the baselines and the GNN used Successive Halving \cite{MLSYS2020_a06f20b3} as a pruning strategy. Successive Halving is a hyperparameter optimisation strategy that tries to identify optimal hyperparameter settings by aggressively pruning underperforming trials. It evaluates different trials at different "rungs," where each run starts with minimal resource allocation, such as epochs. At each subsequent rung, only the highest performing trials, determined by a reduction factor $\eta$, are allowed to progress and use more resources. Allowing resources to be redirected to the most promising trials.

For the sampler, we have used Optuna's TPESampler. TPESampler is the implementation of the Tree-Structured Parzen Estimator (TPE) \cite{NIPS2011_86e8f7ab}. TPE is a Bayesian optimisation method that uses Sequential Model-Based Optimisation instead of Gaussian Processes. Instead of modelling how the hyperparameters affect the score, TPE models the probability density of the hyperparameters that produces the best score $l(x)$ and the density of the hyperparameters that produces the worse score $g(x)$. TPE tries to maximise the ratio between the two densities by identifying the highest Expected Improvement (EI).

The Optuna TPESampler and SuccessiveHalving settings can be found in Table \ref{tab:optuna_config}. The hyperparameter search space for the GNN and the baselines can be found in our repository as well as the selected model hyperparameters. For our optimisation, we used the compound metric derived during the validation step in the training. The compound metric is a weighted metric that combines different metrics to create a single metric for which Optuna will focus on optimising. This metric incorporates important information about different aspects of the model. The compound metric can be seen in Equation \ref{CompScore}.

\begin{equation}
\label{CompScore}
    CompScore = 0.25 * Acc + 0.25 * AUROC + 0.5 * (1-MAE)
\end{equation}

\begin{table}[htbp]
\centering
\caption{Hyperparameter Optimization Configuration}
\label{tab:optuna_config}
\begin{tabular}{@{}lll@{}}
\toprule
\textbf{Component} & \textbf{Parameter} & \textbf{Value} \\ \midrule
\textbf{Sampler (TPE)} & Seed & 42 \\
 & Startup Trials ($n_{startup}$) & 10 \\
 & EI Candidates ($n_{ei}$) & 24 \\
 & Multivariate & True \\
 & Group & True \\ \midrule
\textbf{Pruner (SHA)} & Minimum Resource ($R_{min}$) & 8 \\
 & Reduction Factor ($\eta$) & 3 \\
 & Min. Early Stopping Rate & 0 \\
 & Bootstrap Count & 2 \\ \bottomrule
 \textbf{Training Parameters} & Training Epochs & 75 \\
 & Number of DataLoader Workers & 8 \\
 & Batch Size & 64 \\
 & Pin Memory & True \\
 & Persistent Workers & True \\
 & Window Length & 512 \\
 & Window Stride & 512 \\
 & Gradient Accumulation Steps GNN & 2 \\
 & Gradient Accumulation Steps Baselines & 1 \\\bottomrule
\end{tabular}
\end{table}

\begin{figure*}[!t]
\centering
\includegraphics[width=0.9\linewidth]{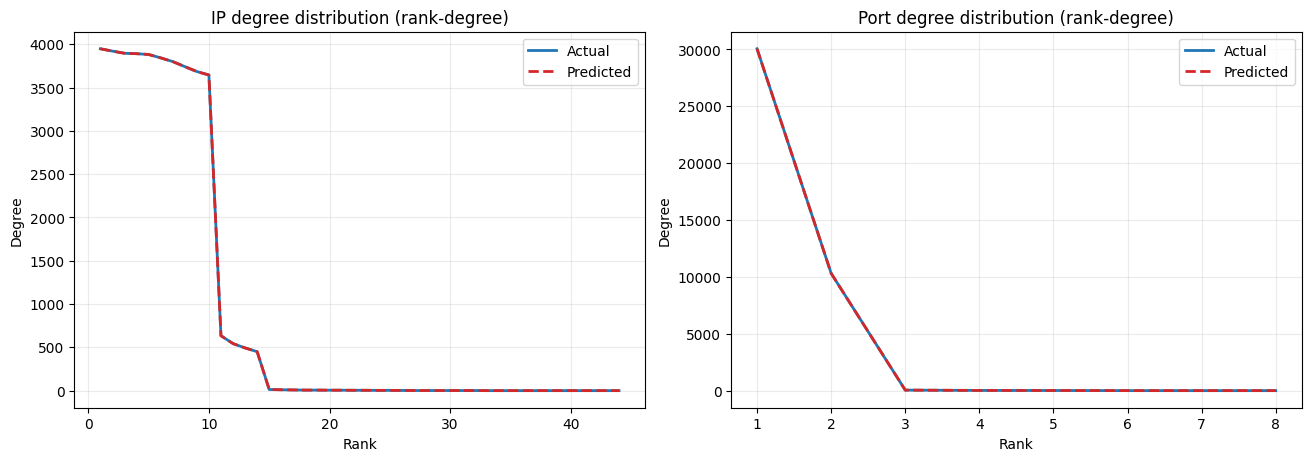}
\caption{The actual ranked connectivity degree of the IP nodes and the forecasted degree by the GNN. Ranked degrees showcase the IPs and Ports that have the highest to the lowest connectivity.}
    \label{fig:ip_degree_ranked}
\end{figure*}

\begin{table*}[t]
\centering
\caption{Model Performance: Feature Reconstruction vs. Structural Prediction}
\label{tab:results_split}
\small
\begin{tabular}{@{}l | cc | ccc@{}}
\toprule
 & \multicolumn{2}{c|}{\textbf{Feature Reconstruction}} & \multicolumn{3}{c}{\textbf{Structural Reconstruction}} \\
\textbf{Model} & \textbf{MAE} ($\downarrow$) & \textbf{MSE} ($\downarrow$) & \textbf{Accuracy} ($\uparrow$) & \textbf{AUROC} ($\uparrow$) & \textbf{Precision} ($\uparrow$) \\ \midrule
GNN & \textbf{0.0647} & 0.0196 & \textbf{87.9\%} & \textbf{95.1\%} & \textbf{87.8\%} \\ \midrule
LSTM & 0.0650 & \textbf{0.0194} & 16.4\% & 41.1\% & 19.2\% \\
TCN & 0.0650 & 0.0195 & 16.4\% & 39.3\% & 15.4\% \\
Transformer & 0.0660 & 0.0196 & 16.2\% & 41.1\% & 15.5\% \\
DLinear & 0.0660 & 0.0197 & 16.4\% & 39.1\% & 15.7\% \\ \bottomrule
\end{tabular}
\end{table*}

\subsection{Results}

Regarding the metrics used in this paper, we chose to use MSE and MAE for the numerical feature reconstruction, and for the structural data, we are using Macro metrics for Accuracy, AUROC, and Precision. For the evaluation, we are choosing the Baseline Models and the GNN that have achieved the highest compound score and the test split of the dataset. The hyperparameters used and dataset splits can be found in our code repository.

Looking at the results in Table \ref{tab:results_split}, we can see that the proposed GNN outperforms all baselines when it comes to reconstructing the IP and Port information. This is because of the GNN's ability to model structural dependencies in the Graphs. This allows for better modelling of the IP and Port connections, which is necessary to identify where different NetFlow flows occur. In the case of the baselines, we can clearly see that no baseline is able to identify how each NetFlow is connected to the correct IP address and Port, with their Accuracy being at 16\% for the baselines and 87.5\% for the GNN. The same can be seen in the case of AUROC and Precision metrics where the GNN clearly outperforms the baseline models, again demonstrating the advantage of graph representation for forecasting tasks. This can also be seen in Figure \ref{fig:ip_degree_ranked}, which shows the degree of connectivity of the IP and Port nodes. The nodes are ranked from the most connected to the least connected and show how many edges originate from them. Our proposed GNN can be seen to perform exceptionally well by matching the degree connectivity of the actual nodes.

Looking at the reconstruction of features, we can see a similar approach; overall, the GNN is very closely correlated to the LSTM baseline as seen in Table \ref{tab:results_split}. LSTM only slightly outperforms the GNN at the MSE but on MAE GNN remains the best approach. This indicates that the GNN seems to be performing exceptionally well in the majority of the data, yet it struggles when it comes to outliers, as seen by the MSE error, whereas LSTM is performing better at the outliers, but tends to be less accurate at the majority of the data.

\section{Conclusion and Future Work}

The proposed GNN showcases the capabilities of using graphs to model NetFlows at a fine-grained level compared to previous research that focused on aggregated metrics. Our model is able to outperform existing baselines and showcases that GNNs can successfully model NetFlow connectivity and their features.

Our future work will focus on improving training and evaluating performance on a larger set of datasets, supporting variable graph sizes, which is necessary to better model the non-static nature of network traffic, and controlled ablations of the graph representation, masking, enhanced dot-product decoder, and edge dropping.

\section*{Acknowledgements}

This work was supported by Grant PCI2023-145974-2 funded by MICIU/AEI/10.13039/501100011033 and cofunded by the European Union (GRAPHS4SEC project). This work is also part of the I+D+i project titled BLOSSOMS, grant PID2024-158530OB-I00, funded by MICIU/AEI/10.13039/501100011033/ and by ERDF/EU. This work is also supported by the Catalan Institution for Research and Advanced Studies (ICREA). OpenAI Codex was for debugging and development. It served mainly as a search engine for what was causing the bugs and as a recommendation engine to check which libraries and their methods are the most suitable and optimal during development.

\balance
\bibliographystyle{ieeetr}
\bibliography{bibliography}

\end{document}